\documentclass[a4paper,twoside]{article}
\usepackage[utf8]{inputenc}  
\usepackage{epsfig}
\usepackage{subcaption}
\usepackage{calc}
\usepackage{amssymb}
\usepackage{amstext}
\usepackage{amsmath}
\usepackage{amsthm}
\usepackage{multicol}
\usepackage{pslatex}
\usepackage{apalike}
\usepackage{SCITEPRESS}     

\begin{document}

\title{Persuasion Meets AI: Ethical Considerations for the Design of Social Engineering Countermeasures}

\author{\authorname{Nicolás E. Díaz Ferreyra\sup{1}\orcidAuthor{0000-0001-6304-771X}, Esma Aïmeur\sup{2}, \\Hicham Hage\sup{3}, Maritta Heisel\sup{1} and Catherine García van Hoogstraten\sup{4}}
\affiliation{\sup{1}University of Duisburg-Essen, Department of Computer Science and Applied Cognitive Science, Germany}
\affiliation{\sup{2}University of Montr\'{e}al, Department of Computer Science and Operations Research (DIRO), Canada}
\affiliation{\sup{3}Notre Dame University - Louaize, Computer Science Department, Lebanon}
\affiliation{\sup{4}Booking.com, the Netherlands}
\email{nicolas.diaz-ferreyra@uni-due.de, aimeur@iro.umontreal.ca, hhage@ndu.edu.lb, maritta.heisel@uni-due.de, catherine.garcia@booking.com}}

\keywords{social engineering, digital nudging, privacy awareness, risk management, self-disclosure, AI, ethics}

\abstract{Privacy in Social Network Sites (SNSs) like Facebook or Instagram is closely related to people's self-disclosure decisions and their ability to foresee the consequences of sharing personal information with large and diverse audiences. Nonetheless, online privacy decisions are often based on spurious risk judgements that make people liable to reveal sensitive data to untrusted recipients and become victims of social engineering attacks. Artificial Intelligence (AI) in combination with persuasive mechanisms like nudging is a promising approach for promoting preventative privacy behaviour among the users of SNSs. Nevertheless, combining behavioural interventions with high levels of personalization can be a potential threat to people's agency and autonomy even when applied to the design of social engineering countermeasures. This paper elaborates on the ethical challenges that nudging mechanisms can introduce to the development of AI-based countermeasures, particularly to those addressing unsafe self-disclosure practices in SNSs. Overall, it endorses the elaboration of personalized risk awareness solutions as i) an ethical approach to counteract social engineering, and ii) as an effective means for promoting reflective privacy decisions.}


\onecolumn \maketitle \normalsize \setcounter{footnote}{0} \vfill
\section{Introduction}

Social Network Sites (SNSs) like Instagram and Twitter have changed radically the way people create and maintain interpersonal relationships \cite{penni2017future}. One of the major attractiveness of such platforms is their broadcasting affordances which allow users to connect seamlessly with large and diverse audiences within a few seconds. However, while these platforms effectively contribute to maximizing people's social capital, they also introduce major challenges to their privacy. Particularly, users of SNSs are highly exposed to social engineering attacks since they often share personal information with people regardless of their doubtful trustworthiness \cite{wang2011regretted,boyd2010social}.

Online deception is an attack vector that is frequently used by social engineers to approach and manipulate users of SNSs \cite{tsikerdekis2014online}. For instance, deceivers often impersonate trustworthy entities using fake profiles to gain their victims' trust, and persuade them to reveal sensitive information (e.g. their log-in credentials) or perform hazardous actions that would compromise their security (e.g. installing Malware) \cite{hage2020understanding}. Hence, counteracting social engineering attacks relies (to a large extent) on the users' capacity of foreseeing the potential negative consequences of their actions and modify their behaviour, accordingly \cite{aimeur2019manipulation}. However, this is difficult for average users who lack the knowledge and skills necessary to ensure the protection of their privacy \cite{masur2019privacy}. Moreover, people -in general- regret having shared their personal information only after being victims of a social engineering attack \cite{wang2011regretted}.

Privacy scholars have proposed a wide range of Artificial-Intelligence-Based (AI-based) approaches that aim at generating awareness among people \cite{de2018privacy,petkos2015pscore,diaz2020preventative,briscoe2014cues}. Particularly, AI in combination with persuasive mechanisms has gained popularity due to their capacity for nudging users' behaviour towards safer privacy practices \cite{acquisti2017nudges}. However, these technologies are often looked askance since there is a fine line between persuasion, coercion, and manipulation \cite{renaud2018ethical}. For instance, users might be encouraged to enable their phone's location services with the excuse of increasing their safety when, in fact, the main objective is monitoring their movements. Hence, ethical principles must be well defined and followed for safeguarding people's agency, autonomy, and welfare.

This work elaborates on the ethical challenges associated with the use of persuasion in the design of AI-based countermeasures (i.e. technical solutions for preventing unsafe self-disclosure practices). Particularly, it analyses the different factors that influence online privacy decisions and the importance of behavioural interventions for promoting preventative privacy practices in SNSs. Furthermore, it elaborates on the ethical issues that the use of persuasive means may introduce when used in combination with AI technologies. Based on our findings, we endorse the elaboration of personalized risk awareness mechanisms as an ethical approach to social engineering countermeasures. In line with this, challenges for regulating the development and impact of such countermeasures are evaluated and summarized.

The rest of this paper is organized as follows. Section~\ref{deception} discusses the particular challenges that SNSs introduce in terms of privacy decision-making. Section~\ref{countermeasures} elaborates on the use of privacy nudges in combination with AI for designing effective social engineering countermeasures. Moreover, ethical challenges related to the use of persuasion in cybersecurity are presented and illustrated in this section. Next, Section \ref{approaches} analyses the role of risk cues in users' privacy decisions and their importance for the design of preventative technologies. Finally, the conclusions of this work are presented in Section~\ref{conclusion}.

\section{Online Deception in Social Media} \label{deception}

Nowadays, SNSs offer a wide range of affordances (e.g. instant messaging, posts, or stories) which allow people to create and exchange media content with large and diverse audiences. In such a context, privacy as a human practice (i.e. as a decision-making process) acquires high importance since individuals are prone to disclose large amounts of private information inside these platforms \cite{boyd2010social}. Consequently, preserving users' contextual integrity depends to a wide extent on their individual behaviour, and not so much on the security mechanisms of the platform (e.g. firewalls or communication protocols) \cite{albladi2016vulnerability}. In general, disclosing personal information to others is key for the development and strengthening of social relationships, as it directly contributes to building trust and credibility among individuals. However, unlike in the real world, people in SNSs tend to reveal their personal data prematurely without reflecting much on the potential negative effects \cite{aimeur2018scourge}. On one hand, such spurious behaviour can be grounded on users' ignorance and overconfidence \cite{howah2019we}. Nevertheless, people often rely on lax privacy settings and assume their online peers as trusted, which increases significantly the chances of being victims of a malicious user. Therefore, individuals are prone to experience unwanted incidents like cyber-bullying, reputation damage, or identity theft after sharing their personal information in online platforms \cite{wang2013privacy}.

Overall, SNSs have become a gateway for accessing large amounts of personal information and, consequently, a target for social engineering attacks. On one hand, this is because people are more liable to reveal personal information online than in a traditional offline context. However, there is also a growing trend in cyber-attacks to focus more on human vulnerabilities instead on flaws in software or hardware \cite{krombholz2015advanced}. Moreover, it is estimated that around 3\% of Malware attacks exploit technical lapses while the remaining 97\% target the users using social engineering\footnote{``Estimates of the number of social engineering based cyber-attacks into private or government organization,'' DOGANA H2020 Project. Accessed July 24, 2020. https://bit.ly/2k5VKmP}. Basically, social engineers employ online deception as a strategy to gain trust and manipulate their victims. Particularly, ``deceivers hide their harmful intentions and mislead other users to reveal their credentials (i.e. accounts and passwords) or perform hazardous actions (e.g. install Malware)'' \cite{aimeur2019manipulation}. For instance, they often approach users through fake SNSs accounts and instigate them to install malicious software on their computers. For this, deceivers exploit users' motivations and cognitive biases such as altruism or moral gain in combination with incentive strategies to mislead them, accordingly \cite{bullee2018anatomy}. Particularly, the use of fake links to cash prizes or fake surveys on behalf of trustworthy entities can serve as incentives and, thereby, as deceptive means.

\section{AI-Based Countermeasures} \label{countermeasures}

In general, people struggle to regulate the amount of information they share as they seek the right balance between self-disclosure gratifications and privacy risks. Moreover, an objective evaluation of such risks demands a high cognitive effort which is often affected by personal characteristics, emotions, or missing knowledge \cite{kramer2019mastering}. Hence, there is a call for technological countermeasures that support users in conducting a more accurate privacy calculus and incentivize the adoption of preventative behaviour. In this section, we discuss the role of AI in the design of such countermeasures especially in combination with persuasive technologies like digital nudges. Furthermore, ethical guidelines for the application of these technologies are presented and analysed.

\subsection{Privacy nudges}

The use of persuasion in social computing applications like blogs, wiki, and recently SNSs has caught the interest of researchers across a wide range of disciplines including computer science and cognitive psychology \cite{vassileva2012motivating}. Additionally, the field of behavioural economics has contributed largely to this topic and nourished several principles of user engagement such as gamification or incentive mechanisms for promoting behavioural change \cite{hamari2013social}. Most recently, the nudge theory and its application for privacy and security purposes have been closely explored and documented within the literature \cite{acquisti2017nudges}. Originally, the term nudge was coined by the Nobel prize winners Richard Thaler and Cass Sunstein and refers to the introduction of small changes in a choice architecture  (i.e. the context within which decisions are made) with the purpose of encouraging a certain user behaviour \cite{weinmann2016digital}, Among its many applications, the nudge concept has been applied in the design of preventative technologies with the aim of guiding users towards safer privacy decisions. For example, \cite{wang2013privacy} designed three nudges for Facebook users consisting of (i) introducing a 30 seconds delay before a message is posted, (ii) displaying visual cues related to the post's audience, and (iii) showing information about the sentiment of the post. These nudges come into play when users are about to post a message on Facebook allowing them to reconsider their disclosures and reflect on the potential privacy consequences. Moreover, nudges have also been designed, developed, and applied for security purposes. This is the case of password meters used to promote stronger passwords \cite{egelman2013does} or the incorporation of visual cues inside Wi-Fi scanners to encourage the use of secure networks \cite{turland2015}.

\subsection{The role of AI}
In general, the instances of privacy nudges described in the current literature rely on a ``one-size-fits-all'' persuasive design. That is, the same behavioural intervention is applied to diverse individuals without acknowledging the personal characteristics or differences among them \cite{warberg2019can}. However, there is an increasing demand for personalized nudges that address nuances in users' privacy goals and regulate their interventions, accordingly \cite{peer2019nudge,barev2019}.

\begin{figure}[h]
\centering
\vspace{1ex}
\includegraphics[width=\linewidth]{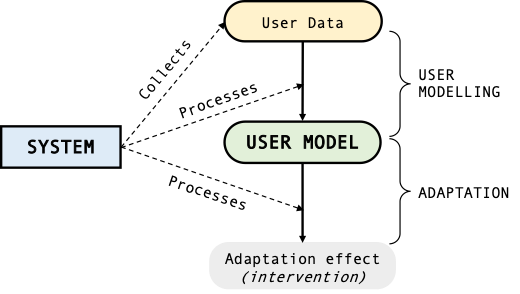}
\vspace{0.5ex}
\caption{The user modelling-adaptation loop \cite{de2017challenges}}
\label{fig:1}
\end{figure}

In essence, the idea of personalized nudges inherently encloses the application of AI techniques and methods for understanding and anticipating the privacy needs of each particular user. For this, it is necessary to define what is commonly known as the ``user model'' of the system. That is, a set of adaptation variables that will guide the personalization process of behavioural interventions \cite{de2017challenges}. For instance, people's privacy attitude has been often proposed as an adaptation means in the design of privacy wizards. Under this approach, users are classified into fundamentalists, pragmatists, or unconcerned, and their privacy policies adjusted to each of these categories \cite{knijnenburg2014information,alekh2018human}. As result, fundamentalists receive strong privacy policies, moderate settings are assigned to pragmatists, and weak ones to unconcerned users. In this case, the user model is said to be explicit since it is generated out information that the users provide before starting to use the system (e.g. in an attitude questionnaire). However, the need for explicit user input can be diminished when implicit models are automatically generated from large data sets \cite{de2017challenges}. Under this approach, the model is automatically obtained out of information that emerges from the interaction between the user and the system. Particularly, information such as likes, clicks, and comments is aggregated into an implicit model that guides the adaptation of the system's interventions (Figure~\ref{fig:1}).

In general, the application of AI to nudging solutions offers the potential of boosting the effectiveness of behavioural interventions. However, such effectiveness comes with a list of drawbacks inherited from the underlying principles of AI technologies. Indeed, as personalization in nudges increases, concerns related to automated decision making and profiling quickly arise along with issues of transparency, fairness, and explainability \cite{susser2019invisible,brundage2018malicious}. Consequently, the user model and adaptation mechanism underlying these nudges should be scrutable in order to prevent inaccurate, unfair, biased, or discriminatory interventions. This would not only improve the system's accountability but would also give insights to the users on how their personal data is being used to promote changes in their behaviour. Over the last years, explainable AI (XAI) has shed light on many of these points and introduced methods for achieving transparent and accountable solutions. One example is the introduction of self-awareness mechanisms that endow deep learning systems with the capability to automatically evaluate their own beliefs and detect potential biases \cite{garigliano2019looking}. However, the combination of AI with persuasion introduces additional challenges related to the impact that these technologies may have on the resulting behaviour of individuals and society in general \cite{sep-ethics-ai,susser2019invisible}. Hence, the definition of ethical guidelines, codes of conduct, and legal provisions are critical for guiding the development process of personalized nudges and for preventing a negative effect on people's well-being.

\subsection{Ethical Guidelines}

Although many have shown excitement about nudges and their applications, others consider this type of persuasive technologies as a potential threat to the users' agency, and autonomy \cite{renaud2018ethical,susser2019invisible}. Particularly, some argue that nudges do not necessarily contribute to users' welfare and could even be used for questionable and unethical purposes. For instance, a mobile application can nudge users to enable their phone's location services with the excuse of improving the experience within the app when, in fact, the main purpose is monitoring their movements. One case alike took place recently in China during the outbreak of the Coronavirus: the Chinese government implemented a system to monitor the virus's expansion and notify citizens in case they need to self-quarantine\footnote{ Paul Mozur, Raymond Zhong and Aaron Krolik ``In Coronavirus Fight, China Gives Citizens a Color Code, With Red Flags'' New York Times, March 1, 2020. Accessed July 24, 2020.}. Such a system generates a personal QR code which is scanned by the police and other authorities to determine whether someone is allowed into subways, malls, and other public spaces. However, although the system encourages people to provide personal information such as location and body temperature on behalf of public safety, experts suggest that this is another attempt by the Chinese government to increase mass surveillance.

\begin{figure}[h]
\centering
\includegraphics[width=0.65\linewidth]{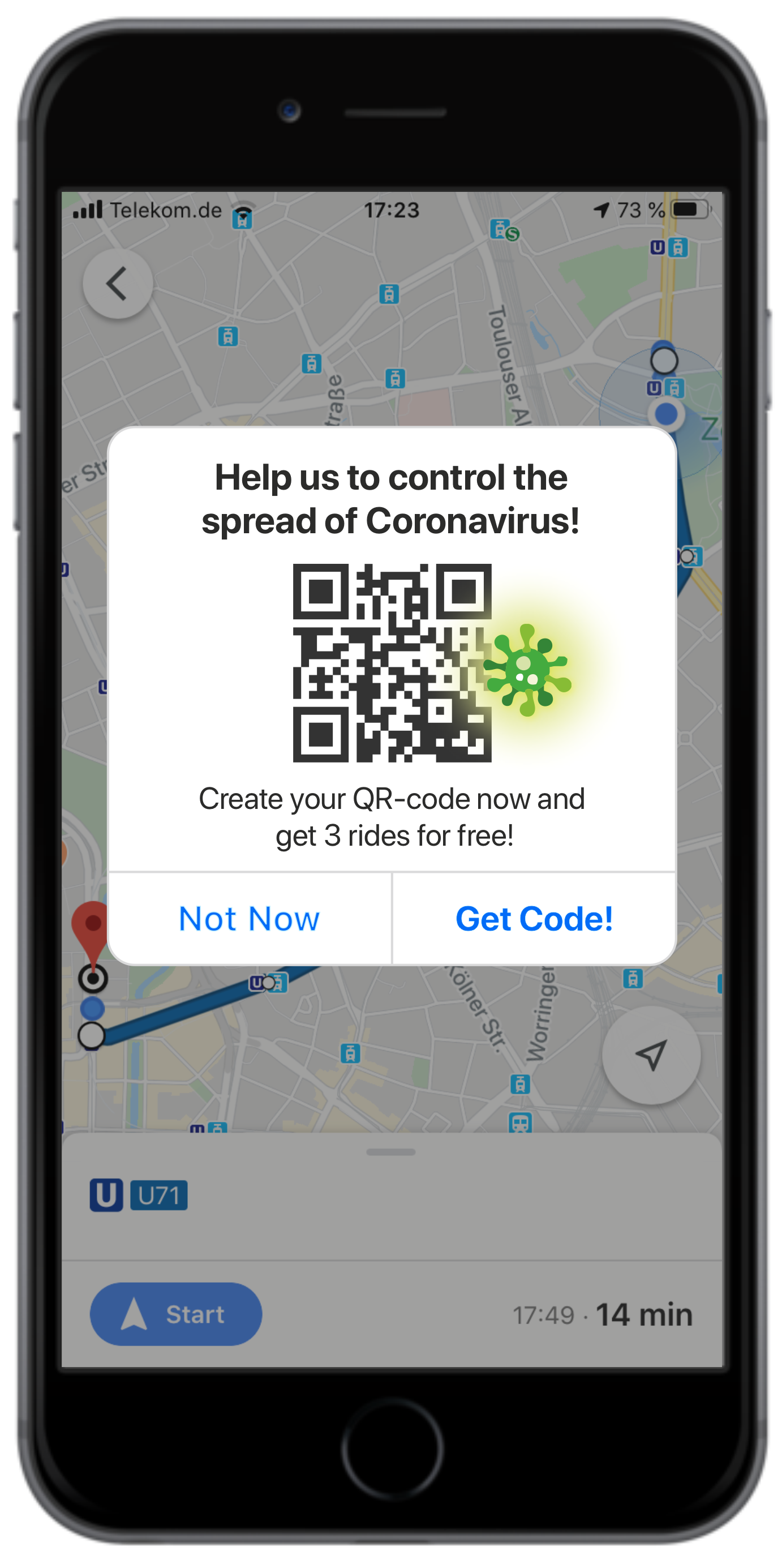}
\caption{Nudge for monitoring the Corovavirus spread}
\label{fig:2}
\vspace{2ex}
\end{figure}

One distinctive aspect of nudges is that they are applied in a decision-making context. That is, they are used to encourage the selection of one alternative over others with the aim of maximizing people's welfare \cite{weinmann2016digital}. In the case of the QR system introduced in China, its adoption was made mandatory by the government. Hence, it does not qualify as a nudge solution since this would normally involve the manipulation of the environment in which a decision is made while preserving users' autonomy and freedom of choice. However, one could imagine a nudge variant of such a system: the QR-code could be included as an optional feature inside a mobile application (e.g. public transport), and its use incentivized through a reward mechanism (e.g. a discount in the next trip) as illustrated in Figure~\ref{fig:2}. This and other instances of choice architectures give rise to ethical questions like ``who should benefit from nudges?'', ``should users be informed of the presence of a nudge?'' and ``how nudges should (not) influence the users?''. In line with this, \cite{renaud2018ethical} elaborated on a set of principles that privacy and security nudges should incorporate into their design to address these and other ethical concerns. Particularly, they introduced check-lists that designers can use to verify if their nudging solutions comply with principles such as justice, beneficence, and respect. For instance, to preserve users' autonomy, designers should ensure that all the original options are made available. This means that, if a nudge attempts to discourage people from installing privacy-invasive apps on their phones, then users should still have the option to install these apps if they wish to. Moreover, users should always be nudged towards behaviours that maximize their welfare rather than the interests of others. That is, choices that enclose a benefit for the designer (if any) should not be prioritized over those who do benefit the user.

\section{Risk-based approaches} \label{approaches}

As discussed, nudges can raise ethical concerns even if they are conceived for seemingly noble purposes like privacy and security. Furthermore, although personalization increases their effectiveness, it can also compromise users' privacy and autonomy. In this section, risk awareness is discussed and presented as a suitable means for developing appropriate nudging solutions. Particularly, it elaborates on how risk cues influence peoples' privacy decisions and how choice architectures may incorporate such cues into their design. Moreover, it discusses state-of-the-art solutions in which risk perception has been introduced as an adaptation variable for personalizing behavioural interventions.

\subsection{The role of risk cues} \label{risk_cues}

Risks are part of our daily life since there is always some uncertainty associated with the decisions we make. Moreover, it is our perception of risk which often helps us to estimate the impact of our actions and influences our behaviour \cite{williams2007does}. However, evaluating a large number of risk factors is often difficult due to the limited cognitive capacity of humans in general. Consequently, people often misjudge the consequences of their actions, behave unseemlily, and suffer unwanted incidents \cite{fischer2017perception}. To avoid this, it is of utmost importance to increase individuals' sense of awareness and so their access to explicit and adequate risk information \cite{Kim2017}. This premise not only applies to decisions that are made in the real world but also in online contexts such as the disclosure of personal information. Particularly, self-disclosure is a practice which is usually performed under uncertainty conditions related to its potential benefits and privacy costs \cite{acquisti2015privacy}. However, average SNSs users find it difficult to perform proper estimations of the risks and benefits of their disclosures and, in turn, replace rational estimations with cognitive heuristics. For example, they often ground their privacy decisions on cues related to the platform's reputation (e.g. it's size) or recognition (e.g. it's market presence), among others \cite{marmion2017cognitive}. All in all, the application of heuristics tends to simplify complex self-disclosure decisions. However, these heuristics can also undermine people's privacy-preserving behaviour since SNSs portray many trust-related cues, yet scarce risk information \cite{marmion2017cognitive}. Furthermore, privacy policies are also devoid of risk cues which, in turn, hinder users' decisions related to consent on data processing activities \cite{de2018consent}. Consequently, even users with high privacy concerns may lack adequate means for conducting a rigorous uncertainty calculus.

\subsection{Risk, persuasion and AI}

In general, risk awareness has a strong influence on people's behaviour and plays a key role in their privacy decisions. Therefore, the presence of risk cues is essential for supporting users in their self-disclosure practices. Under this premise, privacy scholars have introduced nudging solutions that aim to promote changes in people's privacy behaviour using risk information as a persuasive means. For example, \cite{de2018privacy} introduced an approach based on attack-trees and empirical evidence to inform users of SNSs about the privacy risks of using lax privacy settings (e.g. the risks of having a public profile). Similarly, \cite{sanchez2015} developed a method for automatically assessing the sensitivity degree of textual publications in SNSs (e.g. tweets or posts). Such a method takes into consideration the degree of trust a user has in the targeted audience of a message (e.g. family, close friends, acquaintances) when determining its sensitiveness level. Furthermore, this approach is embedded in a system which notifies the users when privacy-critical content is being disclosed and suggests them to either restrict the publication's audience or remove/replace the sensitive terms with less detailed information. Nevertheless, the system ignores nuances in people's privacy goals and does not provide a mechanism for personalizing the interventions.

\begin{figure}[b]
\centering
\vspace{2ex}
\includegraphics[width=\linewidth]{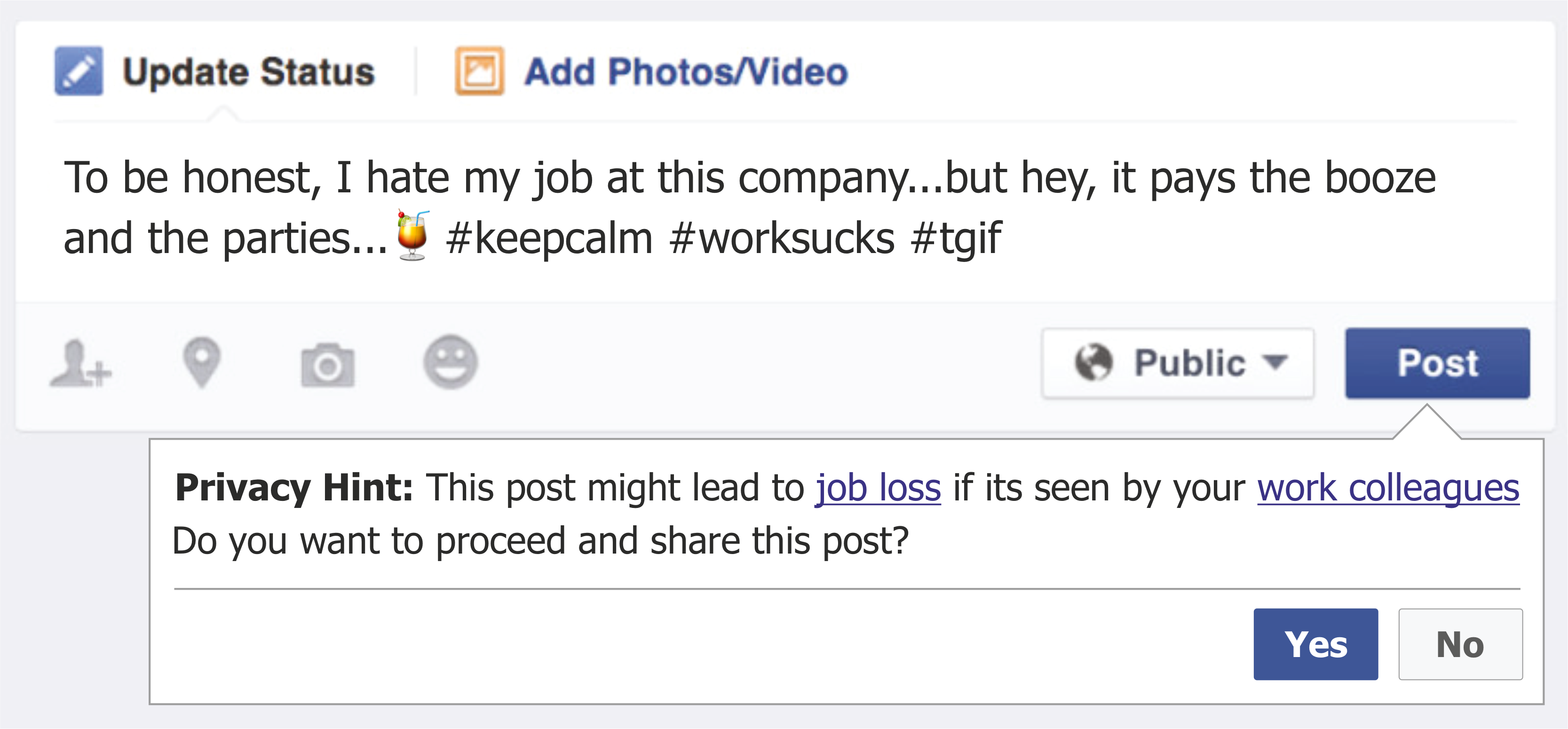}
\caption{Personalized interventions for online-self disclosure \cite{diaz2020preventative}.}
\label{fig:3}
\end{figure}

In order to guide the development of preventative nudges, \cite{diaz2018pst} introduced three design principles. The first one, \textit{adaptivity}, refers to the importance of personalized interventions in creating engagement between the nudge and its users. Particularly, personalization is considered key for engaging individuals in a sustained learning process about good privacy practices. The second one, \textit{viscerality}, highlights the importance of creating a strong and appreciable connection between users and their personal data. This principle is grounded on empirical evidence showing that, in general, users take conscience about the value of their data only after they suffer an unwanted incident (e.g. phishing or financial fraud). Finally, the principle of \textit{sportiveness}, suggests that nudging solutions to cybersecurity should recommend countermeasures or coping strategies that users can put into practice to safeguard their privacy. When it comes to sportiveness, the authors suggest that many of the current privacy-enhancing technologies such as access-control lists and two-step verification would qualify as countermeasures and that the role of nudges is to motivate their adoption. On the other hand, they also suggest that adaptation and viscerality can be achieved by defining a user model which reflects individuals' risk perception \cite{diaz2020preventative}. Particularly, they introduced a user model consisting of a risk threshold which is updated as behavioural interventions are accepted or ignored by the end-user. By doing so, the nudge adapts to the individual privacy goals of the users and increases the effectiveness of its interventions. That approach was put into practice in the design of preventative technologies for SNSs as depicted in Figure~\ref{fig:3}. Particularly, this approach uses empirical evidence on regrettable self-disclosure experiences to elaborate risk patterns and shape behavioural interventions. Such risk-based interventions aim to encourage the use of friend lists for controlling the audience of textual publications.

\section{Discussion and Conclusion} \label{conclusion}

Overall, social interaction across SNSs demands making privacy decisions on a frequent basis. However, online self-disclosure, as well as the evolving and ongoing nature of privacy choices, seem to be out of the scope of privacy regulations. Instead, data protection frameworks tend to focus more on issues related to consent and overlook (sometimes deliberately) the importance of providing the necessary means for performing an adequate privacy calculus. Consequently, service providers limit themselves to the definition of instruments for obtaining consent (e.g. privacy policies) leaving the rational estimations of privacy risks to the individual discretion of the users. However, as discussed throughout this paper, such estimations are often impaired by cognitive and motivational biases which tend to outweigh anticipated benefits over potential risks. Hence, users are prone to experience regret after disclosing personal information in SNSs due to false estimations and optimistic biases. Furthermore, because of such spurious estimations, people might end up sharing their private information with untrusted audiences and increase their chances of suffering social engineering attacks. Thus, the incorporation of awareness mechanisms is of utmost importance for supporting users' self-disclosure decisions and mitigating the likelihood of unwanted incidents.

At their core, social engineering countermeasures require promoting behavioural changes among the users of SNSs. Hence, nudging techniques are of great value for the design of technical solutions that could guide individuals towards safer privacy decisions. Furthermore, AI-based approaches can improve the effectiveness of such solutions by endowing them with adaptation and personalization features that address the individual goals and concerns of the users. However, despite its promising effects, the combination of AI together with persuasive means can result in unethical technological designs. In principle, this is because nudges can mislead people towards a behaviour which is not necessarily beneficial for them. In this case, ethical guidelines are quite clear since they stress the importance of designing choice architectures that are transparent and tend to maximize people's welfare. However, the question that still remains unclear is whether platforms should keep self-regulating the design of these technologies, or if public actors should harness the application of ethical standards in order to safeguard individuals' agency and autonomy. Either way, conducting an adequate \textit{social welfare impact assessment} of persuasive AI technologies is crucial to determine their effects (positive or negative) on human behaviour at large.

Another controversial point is that users are not always aware of the presence of a nudge since persuasive means target primarily people's automatic and subconscious processing system. However, ethical approaches should allow people to explicitly recognize the presence of the nudge and the influence it is aiming to excerpt. Hence, choice architectures should, when possible, introduce mechanisms that target individuals' reflective reasoning in order to avoid potential manipulation effects. As discussed in Section~\ref{risk_cues}, social engineering countermeasures can achieve this by incorporating risk information and cues in their design. However, even risk-related information can be subject to manipulation if not framed accordingly. That is, when high-risk events are portrayed as low-risk situations and vice-versa. Furthermore, biases can also be introduced if the likelihood and consequence levels of unwanted incidents are not properly estimated and quantified. Hence, guidelines for the correct estimation of risks together with ethical approaches for their communication should be further investigated, introduced, and guaranteed.

\section*{\uppercase{Acknowledgements}}

This work was partially supported the H2020 European Project No. 787034 ``PDP4E: Privacy and Data Protection Methods for Engineering'' and Canada's Natural Sciences and Engineering Research Council (NSERC).

\bibliographystyle{apalike}

{\small 

\begin{thebibliography}{}

\bibitem[Acquisti et~al., 2017]{acquisti2017nudges}
Acquisti, A., Adjerid, I., Balebako, R., Brandimarte, L., Cranor, L.~F.,
  Komanduri, S., Leon, P.~G., Sadeh, N., Schaub, F., Sleeper, M., Wang, Y., and
  Wilson, S. (2017).
\newblock {Nudges for Privacy and Security: Understanding and Assisting Users'
  Choices Online}.
\newblock {\em ACM Computing Surveys (CSUR)}, 50(3):44.

\bibitem[Acquisti et~al., 2015]{acquisti2015privacy}
Acquisti, A., Brandimarte, L., and Loewenstein, G. (2015).
\newblock Privacy and human behavior in the age of information.
\newblock {\em Science}, 347(6221):509--514.

\bibitem[A\"{i}meur et~al., 2019]{aimeur2019manipulation}
A\"{i}meur, E., Diaz~Ferreyra, N.~E., and Hage, H. (2019).
\newblock {Manipulation and Malicious Personalization: Exploring the
  Self-Disclosure Biases Exploited by Deceptive Attackers on Social Media}.
\newblock {\em Frontiers in Artificial Intelligence}, 2:26.

\bibitem[A\"{i}meur et~al., 2018]{aimeur2018scourge}
A\"{i}meur, E., Hage, H., and Amri, S. (2018).
\newblock {The Scourge of Online Deception in Social Networks}.
\newblock In Arabnia, H.~R., Deligiannidis, L., Tinetti, F.~G., and Tran,
  Q.-N., editors, {\em Proceedings of the 2018 Annual Conference on
  Computational Science and Computational Intelligence (CSCI' 18)}, pages
  1266--1271. IEEE Computer Society.

\bibitem[Albladi and Weir, 2016]{albladi2016vulnerability}
Albladi, S. and Weir, G. R.~S. (2016).
\newblock {Vulnerability to Social Engineering in Social Networks: A Proposed
  User-Centric Framework}.
\newblock In {\em 2016 IEEE International Conference on Cybercrime and Computer
  Forensic (ICCCF)}, pages 1--6. IEEE.

\bibitem[Alekh, 2018]{alekh2018human}
Alekh, S. (2018).
\newblock {Human Aspects and Perception of Privacy in Relation to
  Personalization}.
\newblock {\em arXiv preprint arXiv:1805.08280}.

\bibitem[Barev and Janson, 2019]{barev2019}
Barev, T.~J. and Janson, A. (2019).
\newblock {Towards an Integrative Understanding of Privacy Nudging - Systematic
  Review and Research Agenda}.
\newblock In {\em 18th Annual Pre-ICIS Workshop on HCI Research in MIS (ICIS)}.

\bibitem[boyd, 2010]{boyd2010social}
boyd, d. (2010).
\newblock {\em {Social Network Sites as Networked Publics: Affordances,
  Dynamics, and Implications}}, pages 47--66.
\newblock Routledge.

\bibitem[Briscoe et~al., 2014]{briscoe2014cues}
Briscoe, E.~J., Appling, D.~S., and Hayes, H. (2014).
\newblock {Cues to Deception in Social Media Communications}.
\newblock In {\em 47th Annual Hawaii International Conference on System
  Sciences (HICSS)}, pages 1435--1443. IEEE.

\bibitem[Brundage et~al., 2018]{brundage2018malicious}
Brundage, M., Avin, S., Clark, J., Toner, H., Eckersley, P., Garfinkel, B.,
  Dafoe, A., Scharre, P., Zeitzoff, T., Filar, B., et~al. (2018).
\newblock {The Malicious Use of Artificial Intelligence: Forecasting,
  Prevention, and Mitigation}.
\newblock {\em arXiv preprint arXiv:1802.07228}.

\bibitem[Bull{\'e}e et~al., 2018]{bullee2018anatomy}
Bull{\'e}e, J.-W.~H., Montoya, L., Pieters, W., Junger, M., and Hartel, P.
  (2018).
\newblock {On the anatomy of social engineering attacks - A literature-based
  dissection of successful attacks}.
\newblock {\em Journal of Investigative Psychology and Offender Profiling},
  15(1):20--45.

\bibitem[De and Imine, 2019]{de2018consent}
De, S.~J. and Imine, A. (2019).
\newblock {On Consent in Online Social Networks: Privacy Impacts and Research
  Directions (Short Paper)}.
\newblock In Zemmari, A., Mosbah, M., Cuppens-Boulahia, N., and Cuppens, F.,
  editors, {\em Risks and Security of Internet and Systems}, pages 128--135.
  Springer International Publishing.

\bibitem[De and Le~M{\'e}tayer, 2018]{de2018privacy}
De, S.~J. and Le~M{\'e}tayer, D. (2018).
\newblock {Privacy Risk Analysis to Enable Informed Privacy Settings}.
\newblock In {\em 2018 IEEE European Symposium on Security and Privacy
  Workshops (EuroS\&PW)}, pages 95--102. IEEE.

\bibitem[De~Bra, 2017]{de2017challenges}
De~Bra, P. (2017).
\newblock {Challenges in User Modeling and Personalization}.
\newblock {\em IEEE Intelligent Systems}, 32(5):76--80.

\bibitem[D{\'\i}az~Ferreyra et~al., 2018]{diaz2018pst}
D{\'\i}az~Ferreyra, N., Meis, R., and Heisel, M. (2018).
\newblock {At Your Own Risk: Shaping Privacy Heuristics for Online
  Self-disclosure}.
\newblock In {\em 2018 16th Annual Conference on Privacy, Security and Trust
  (PST)}, pages 1--10.

\bibitem[D{\'\i}az~Ferreyra et~al., 2020]{diaz2020preventative}
D{\'\i}az~Ferreyra, N.~E., Kroll, T., A{\"\i}meur, E., Stieglitz, S., and
  Heisel, M. (2020).
\newblock {Preventative Nudges: Introducing Risk Cues for Supporting Online
  Self-Disclosure Decisions}.
\newblock {\em Information}, 11(8):399.

\bibitem[Egelman et~al., 2013]{egelman2013does}
Egelman, S., Sotirakopoulos, A., Muslukhov, I., Beznosov, K., and Herley, C.
  (2013).
\newblock {Does my password go up to eleven? The impact of password meters on
  password selection}.
\newblock In {\em Proceedings of the SIGCHI Conference on Human Factors in
  Computing Systems}, pages 2379--2388.

\bibitem[Fischer, 2017]{fischer2017perception}
Fischer, A. R.~H. (2017).
\newblock {Perception of Product Risks}.
\newblock In Emilien, G., Weitkunat, R., and L{\"u}dicke, F., editors, {\em
  {Consumer Perception of Product Risks and Benefits}}, pages 175--190.
  Springer International Publishing, Cham.

\bibitem[Garigliano and Mich, 2019]{garigliano2019looking}
Garigliano, R. and Mich, L. (2019).
\newblock {Looking Inside the Black Box: Core Semantics Towards Accountability
  of Artificial Intelligence}.
\newblock In {\em From Software Engineering to Formal Methods and Tools, and
  Back}, pages 250--266. Springer.

\bibitem[Hage et~al., 2020]{hage2020understanding}
Hage, H., A\"{i}meur, E., and Guedidi, A. (2020).
\newblock Understanding the landscape of online deception.
\newblock In {\em Navigating Fake News, Alternative Facts, and Misinformation
  in a Post-Truth World}, pages 290--317. IGI Global.

\bibitem[Hamari and Koivisto, 2013]{hamari2013social}
Hamari, J. and Koivisto, J. (2013).
\newblock {Social Motivations To Use Gamification: An Empirical Study Of
  Gamifying Exercise}.
\newblock In {\em Proceedings of the 21st European Conference on Information
  Systems (ECIS)}.

\bibitem[Howah and Chugh, 2019]{howah2019we}
Howah, K. and Chugh, R. (2019).
\newblock {Do We Trust the Internet?: Ignorance and Overconfidence in
  Downloading and Installing Potentially Spyware-Infected Software}.
\newblock {\em Journal of Global Information Management (JGIM)}, 27(3):87--100.

\bibitem[Kim, 2017]{Kim2017}
Kim, H.~K. (2017).
\newblock {Risk Communication}.
\newblock In Emilien, G., Weitkunat, R., and L{\"u}dicke, F., editors, {\em
  {Consumer Perception of Product Risks and Benefits}}, pages 125--149.
  Springer International Publishing, Cham.

\bibitem[Knijnenburg, 2014]{knijnenburg2014information}
Knijnenburg, B.~P. (2014).
\newblock {Information Disclosure Profiles for Segmentation and
  Recommendation}.
\newblock In {\em Symposium on Usable Privacy and Security (SOUPS)}.

\bibitem[Kr{\"a}mer and Sch{\"a}wel, 2020]{kramer2019mastering}
Kr{\"a}mer, N.~C. and Sch{\"a}wel, J. (2020).
\newblock Mastering the challenge of balancing self-disclosure and privacy in
  social media.
\newblock {\em Current Opinion in Psychology}, 31.

\bibitem[Krombholz et~al., 2015]{krombholz2015advanced}
Krombholz, K., Hobel, H., Huber, M., and Weippl, E. (2015).
\newblock Advanced social engineering attacks.
\newblock {\em Journal of Information Security and applications}, 22:113--122.

\bibitem[Marmion et~al., 2017]{marmion2017cognitive}
Marmion, V., Bishop, F., Millard, D.~E., and Stevenage, S.~V. (2017).
\newblock {The Cognitive Heuristics Behind Disclosure Decisions}.
\newblock In {\em International Conference on Social Informatics}, pages
  591--607. Springer.

\bibitem[Masur, 2019]{masur2019privacy}
Masur, P.~K. (2019).
\newblock Privacy and self-disclosure in the age of information.
\newblock In {\em Situational Privacy and Self-Disclosure}, pages 105--129.
  Springer.

\bibitem[Müller, 2020]{sep-ethics-ai}
Müller, V.~C. (2020).
\newblock {Ethics of Artificial Intelligence and Robotics}.
\newblock In Zalta, E.~N., editor, {\em The {Stanford} Encyclopedia of
  Philosophy}. Metaphysics Research Lab, Stanford University, fall 2020
  edition.

\bibitem[Peer et~al., 2020]{peer2019nudge}
Peer, E., Egelman, S., Harbach, M., Malkin, N., Mathur, A., and Frik, A.
  (2020).
\newblock {Nudge Me Right: Personalizing Online Nudges to People's
  Decision-Making Styles}.
\newblock {\em Computers in Human Behavior}, 109:106347.

\bibitem[Penni, 2017]{penni2017future}
Penni, J. (2017).
\newblock {The Future of Online Social Networks (OSN): A Measurement Analysis
  Using Social Media Tools and Application}.
\newblock {\em Telematics and Informatics}, 34(5):498--517.

\bibitem[Petkos et~al., 2015]{petkos2015pscore}
Petkos, G., Papadopoulos, S., and Kompatsiaris, Y. (2015).
\newblock {PScore: A Framework for Enhancing Privacy Awareness in Online Social
  Networks}.
\newblock In {\em 2015 10th International Conference on Availability,
  Reliability and Security}, pages 592--600. IEEE.

\bibitem[Renaud and Zimmermann, 2018]{renaud2018ethical}
Renaud, K. and Zimmermann, V. (2018).
\newblock Ethical guidelines for nudging in information security \& privacy.
\newblock {\em International Journal of Human-Computer Studies}, 120:22--35.

\bibitem[Sanchez and Viejo, 2015]{sanchez2015}
Sanchez, D. and Viejo, A. (2015).
\newblock {Privacy Risk Assessment of Textual Publications in Social Networks}.
\newblock In {\em Proceedings of the International Conference on Agents and
  Artificial Intelligence - Volume 1}, ICAART 2015, pages 236--241, Setubal,
  PRT. SCITEPRESS - Science and Technology Publications, Lda.

\bibitem[Susser, 2019]{susser2019invisible}
Susser, D. (2019).
\newblock {Invisible Influence: Artificial Intelligence and the Ethics of
  Adaptive Choice Architectures}.
\newblock In {\em Proceedings of the 2019 AAAI/ACM Conference on AI, Ethics,
  and Society}, pages 403--408.

\bibitem[Tsikerdekis and Zeadally, 2014]{tsikerdekis2014online}
Tsikerdekis, M. and Zeadally, S. (2014).
\newblock {Online Deception in Social Media}.
\newblock {\em Communications of the ACM}, 57(9):72.

\bibitem[Turland et~al., 2015]{turland2015}
Turland, J., Coventry, L., Jeske, D., Briggs, P., and van Moorsel, A. (2015).
\newblock {Nudging towards Security: Developing an Application for Wireless
  Network Selection for Android Phones}.
\newblock In {\em Proceedings of the 2015 British HCI Conference}, British HCI
  '15, pages 193--201, New York, NY, USA. Association for Computing Machinery.

\bibitem[Vassileva, 2012]{vassileva2012motivating}
Vassileva, J. (2012).
\newblock Motivating participation in social computing applications: a user
  modeling perspective.
\newblock {\em User Modeling and User-Adapted Interaction}, 22(1-2):177--201.

\bibitem[Wang et~al., 2013]{wang2013privacy}
Wang, Y., Leon, P.~G., Scott, K., Chen, X., Acquisti, A., and Cranor, L.~F.
  (2013).
\newblock {Privacy Nudges for Social Media: An Exploratory Facebook Study}.
\newblock In {\em Proceedings of the 22nd International Conference on World
  Wide Web}, pages 763--770.

\bibitem[Wang et~al., 2011]{wang2011regretted}
Wang, Y., Norcie, G., Komanduri, S., Acquisti, A., Leon, P.~G., and Cranor,
  L.~F. (2011).
\newblock {``I regretted the minute I pressed share'': A Qualitative Study of
  Regrets on Facebook}.
\newblock In {\em Proceedings of the 7th Symposium on Usable Privacy and
  Security, {SOUPS} 2011}, pages 1--16. ACM.

\bibitem[Warberg et~al., 2019]{warberg2019can}
Warberg, L., Acquisti, A., and Sicker, D. (2019).
\newblock {Can Privacy Nudges be Tailored to Individuals' Decision Making and
  Personality Traits?}
\newblock In {\em {Proceedings of the 18th ACM Workshop on Privacy in the
  Electronic Society}}, WPES'19, pages 175--197. ACM.

\bibitem[Weinmann et~al., 2016]{weinmann2016digital}
Weinmann, M., Schneider, C., and vom Brocke, J. (2016).
\newblock {Digital Nudging}.
\newblock {\em Business \& Information Systems Engineering}, 58(6):433--436.

\bibitem[Williams and Noyes, 2007]{williams2007does}
Williams, D.~J. and Noyes, J.~M. (2007).
\newblock {How does our perception of risk influence decision-making?
  Implications for the design of risk information}.
\newblock {\em {Theoretical Issues in Ergonomics Science}}, 8(1):1--35.

\end{thebibliography}
}

\end{document}